\documentclass[11pt]{revtex4}
\usepackage{mathtools}
\usepackage{pdflscape}
\usepackage{subfigure}
\usepackage{amsmath,amsfonts}		
\usepackage{amssymb}
\usepackage{color, xcolor}
\usepackage{soul}
\usepackage{ulem}
\usepackage{multirow, makecell}		
\usepackage{threeparttable}			
\usepackage{graphicx}				
\usepackage{changes}
\usepackage{hyperref}

\begin{document}

\title{Polarization modes of gravitational waves in scalar-tensor-Rastall theory}

\author{Yu-Zhi Fan$^{1,2}$}

\author{Xiao-Bin Lai$^{1,2}$}

\author{Yu-Qi Dong$^{1,2}$}

\author{Yu-Xiao Liu$^{1,2}$}
\email{liuyx@lzu.edu.cn, corresponding author}

\affiliation{
	$^{1}$ Lanzhou Center for Theoretical Physics, Key Laboratory of Quantum Theory and Applications of the Ministry of Education, Key Laboratory of Theoretical Physics of Gansu Province, School of Physical Science and Technology, Lanzhou University, Lanzhou 730000, China
	\\
	$^{2}$ Institute of Theoretical Physics $ \& $ Research Center of Gravitation, Lanzhou University, Lanzhou 730000, China\\
}

\begin{abstract}
\textbf{Abstract:} Rastall theory, originally introduced in 1972, suggests a violation of the usual conservation law. We consider two generalizations of Rastall theory: Brans-Dicke-Rastall theory and the newly established scalar-tensor-Rastall theory, the latter being a further generalization of the former. The field equations in these two generalized theories are studied across different parameter spaces, and the polarization modes of gravitational waves, as a key focus, are subsequently investigated. The results show that the polarization modes of gravitational waves in Brans-Dicke-Rastall theory are the same as those in Brans-Dicke theory; specifically, both theories exhibit the plus, cross, and breathing modes. However, in scalar-tensor-Rastall theory, the polarization modes of gravitational waves depend on the parameter space of the theory. Particularly, over a broad range of the parameter space, regardless of some special values of the parameters, it allows only two tensor modes, just as in general relativity, without introducing any additional degrees of freedom. This indicates that Rastall theory offers a novel approach to constructing modified gravity theories that propagate only two tensor degrees of freedom. In the remaining regions of the parameter space, there is also one scalar mode in addition to the two tensor modes. The scalar mode can be either a mixture of the breathing and longitudinal modes or just a pure breathing mode, depending on the parameter space. These results will play a crucial role in constraining the theoretical parameters through future gravitational wave detection projects, such as LISA, Taiji, and TianQin.
\end{abstract}

\maketitle

\section{Introduction}
\label{sec: intro}
\par With a series of gravitational wave (GW) detections by LIGO, Virgo, and KAGRA collaborations~\cite{Abbott1, Abbott2, Abbott3, Abbott4, Abbott5}, GWs have emerged as a new tool for exploring and understanding the universe, alongside electromagnetic waves~\cite{ShiPi, AndreaAddazi, Rong-GenCai, Zhi-ChaoZhao, Wang:2021srv, Zong-KuanGuo, Li:2023plm, shao2023perceiving, Gao:2022hho, Battista:2021rlh}, marking the dawn of the era of GW astronomy. Particularly, the polarization contents of GWs are predicted differently by various gravity theories. This enables us to test candidate gravity theories by comparing the theoretical predictions with observational results in future GW detections.

\par Although general relativity is widely accepted and experimentally supported, it faces some theoretical and observational problems. For instance, challenges such as the singularity of spacetime~\cite{cai2022spacetime}, dark matter and dark energy~\cite{smith1936mass, zwicky1933helvetica, Zwicky:1937zza, Peebles:2002gy}, cosmic expansion~\cite{la1989extended}, and quantization~\cite{tHooft:1974toh, Goroff:1985th, han2007fundamental} are inconsistent with general relativity. In order to overcome these difficulties, various modified gravity theories have been proposed~\cite{clifton2012modified}. Lovelock’s theorem~\cite{Lovelock:1971yv, Lovelock:1972vz} provides several approaches to constructing modified gravity theories that differ from general relativity. A simple way is to introduce an extra field in addition to the metric tensor $ g_{\mu\nu} $ to describe gravity. For example, scalar-tensor (ST) theory~\cite{brans1961mach, PhysRevD.1.3209, bergmann1968comments, Maggiore:1999wm, faraoni2004scalar, banerjee2001cosmic}, Einstein-aether theory~\cite{Einstein-aetherTheory}, tensor-vector-scalar theory~\cite{TeVeSTheory}, and bi-metric theory~\cite{Rosen:1940zza, Rosen:1940zz, rosen1973bi, rosen1975bi} have been built in this way. Particularly, there is an additional scalar field nonminimally coupled to curvature in ST theory. The well-known Brans-Dicke (BD) theory is the simplest ST theory proposed by Brans and Dicke in 1961 with the aim of being compatible with Mach's principle~\cite{brans1961mach}. And Horndeski theory is the most general ST theory with second-order field equations~\cite{HorndeskiTheory}. Another way to construct modified gravity theories is to allow higher derivatives of the metric in field equations. A typical example is $ f(R) $ theory~\cite{Buchdahl:1970ldb, Baibosunov:1990qm, Sotiriou:2008rp, DeFelice:2010aj, Nojiri:2010wj, Cui:2020fiz, Chen:2020zzs}.

\par Besides the above-mentioned modified gravity theories, the theory proposed by Rastall in 1972~\cite{Rastall1972Generalization} has recently obtained a renewed interest~\cite{Moradpour:2017lcq, Visser:2017gpz, Zhong:2022wlw, Darabi:2017coc, Tan:2024url} for its good agreement with cosmological observational data~\cite{Batista:2011nu, Fabris:2012hw}. Rastall theory considers a nonminimal coupling between matter and geometry, which breaks the divergence-free condition in Lovelock’s theorem. In Rastall theory, the energy-momentum tensor satisfies $ \nabla^\mu T_{\mu\nu} = \frac{1-\lambda}{16\pi G}\nabla_\nu R $. Here, the parameter $ \lambda $ measures the deviation from the usual conservation law, $ \nabla^\mu T_{\mu\nu} = 0 $, and general relativity is recovered when $ \lambda = 1 $. Previously, Smalley considered the nonzero divergence of the energy-momentum tensor in the context of BD theory~\cite{Smalley:1974gn}. Further, Brans-Dicke-Rastall (BDR) theory was established in Ref.~\cite{Carames:2014twa}, combining the fundamental principles of BD theory with those of Rastall theory. Apart from these, various generalizations of Rastall theory have been proposed. For instance, a scalar field with self-interaction was introduced into Rastall theory in Refs.~\cite{Bronnikov:2016odv, Fabris:2011wz}. Besides, the modified conservation law was generalized to $ \nabla^\mu T_{\mu\nu} = \frac{1-\lambda}{16\pi G}\nabla_\nu f $, where $ f $ is a function of certain variables, such as $ f = f(R) $~\cite{Lin:2018dgx, Shahidi:2021lxt} or $ f = f(R, T) $~\cite{Lin:2020fue, Mota:2020oiy}. Moreover, the proportionality coefficient in the modified conservation law was no longer considered as a constant, which was discussed in Ref.~\cite{Moradpour:2017shy}. Additionally, the parameter $ \lambda $ was constrained by galaxy-scale strong gravitational lensing and the rotation curves of low surface brightness galaxies in Refs.~\cite{Li:2019jkv, Tang:2019dsk}.

\par On the other hand, the concept of GW polarization modes was first proposed by Einstein in 1916~\cite{einstein1916approximative}. He claimed that, in general relativity, there are only two polarization modes of GWs in the linearized regime: the plus mode and the cross mode. In 1973, Eardley \textit{et al.} showed that in a general four-dimensional metric theory, there can be up to six polarization modes~\cite{Eardley:1973zuo}. These are the plus, cross, vector-$ x $, vector-$ y $, longitudinal, and breathing modes. For modified gravity theories, the possible polarization modes beyond those of general relativity emerge in the presence of additional degrees of freedom, depending on the constraints of field equations on the Riemann tensor. The relationships between the polarization modes of null GWs and the Riemann tensor can be succinctly expressed using the Newman-Penrose formalism~\cite{Newman:1961qr}. It was then extended to the nonnull GWs by Hyun \textit{et al.}~\cite{Hyun:2018pgn}. Additionally, using the gauge invariants in the Bardeen framework is another common strategy to investigate the GW polarization modes~\cite{Bardeen:1980kt, Flanagan:2005yc, Caprini:2018mtu, Alves:2023rxs}. In recent years, the polarizations of GWs have been widely studied in various modified gravity theories, like $ f(R) $~\cite{Liang:2017ahj, Moretti:2019yhs, Gong:2018ybk}, Horndeski~\cite{Hou:2017bqj, Gong:2018ybk}, Palatini-Horndeski~\cite{Dong:2021jtd, Dong:2022cvf}, Einstein-aether~\cite{Gong:2018ybk, Gong:2018cgj}, bumblebee~\cite{Liang:2022hxd}, scalar-tensor-vector~\cite{liu2021gravitational}, $ f(T) $~\cite{Bamba:2013ooa}, dynamical Chern-Simons~\cite{Wagle:2019mdq}, Einstein-dilaton-Gauss-Bonnet~\cite{Wagle:2019mdq}, Rastall~\cite{Moradpour:2017lcq}, general Einstein-vector~\cite{Lai:2024fza}, and generalized Proca~\cite{Dong:2023xyb} theories. And conclusions regarding the polarizations of GWs in metric theory and ST theory were presented in Ref.~\cite{Dong:2023bgt} from a general perspective. In addition, the classification and possible observation of GW polarizations in higher-dimensional spacetime were investigated in Ref.~\cite{LiuYuQiang}, which serves as a probe for the existence of extra dimensions.

\par Although we cannot confirm the presence or absence of extra polarization modes yet, this can be promisingly achieved in more advanced GW detections. Besides the ground-based GW detectors, the pulsar timing array (PTA)~\cite{sazhin1978opportunities, Detweiler:1979wn, foster1990constructing, JGXG202112003} is another powerful tool to detect GWs~\cite{Yi:2023mbm}, such as the North American Nanohertz Observatory for GWs (NANOGrav)~\cite{NANOGrav}, the European PTA \cite{EPTA}, the Parkes PTA~\cite{PPTA}, the Indian PTA~\cite{InPTA}, and the Chinese PTA~\cite{CPTA}. They are more accurate in the nanohertz frequency range of the spectrum. Actually, a preliminary observational indication of the scalar transverse polarization mode of GWs has recently been discovered in the NANOGrav data set by Huang \textit{et al.}~\cite{chen2021non, chen2023search}. Moreover, space-borne GW detection projects such as LISA~\cite{LISA}, Taiji~\cite{Taiji}, and TianQin~\cite{Tianqin}, which are expected to detect the polarization modes of GWs, are currently under construction. Thus, the detection of the polarization modes of GWs is a crucial and hopeful approach to testing candidate gravity theories.

\par In this paper, we will consider a more general ST theory~\cite{clifton2012modified, PhysRevD.1.3209, bergmann1968comments, Maggiore:1999wm} than BD theory~\cite{brans1961mach}, with the coupling function $ \omega(\phi) $ and the potential function $ V(\phi) $. Combining Rastall theory and this ST theory, we will then establish the field equations of scalar-tensor-Rastall (STR) theory as a generalization of BDR theory~\cite{Carames:2014twa}. Next, we will investigate the polarization modes of GWs in BDR theory based on Ref.~\cite{Carames:2014twa} under a Minkowski background. The conclusion is that they are the same as those in BD theory, i.e., the plus, cross, and breathing modes. Subsequently, we will analyze the polarization modes of GWs in STR theory under a Minkowski background. We will demonstrate that the field equations differ in various regions of the parameter space, similar to the behavior of polarizations. Especially over a broad range of the parameter space, there are only two tensor modes, without any additional degrees of freedom. A summary of these results, along with a comparison of the polarizations in other theories, will be provided subsequently.

\par The paper is organized as follows. In Sec.~\ref{Rastall+generalizations}, we first review Rastall theory~\cite{Rastall1972Generalization} and discuss its polarization modes of GWs. Next, we give an overview of BDR theory~\cite{Carames:2014twa}, and further discuss its field equations for different parameters. Finally, we establish STR theory and derive its field equations for different parameters. In Sec.~\ref{GIandPM}, we give a brief introduction to the construction of gauge invariants and illustrate their relationship with the polarization modes. In Sec.~\ref{GWPinBDR}, we study the GW polarizations in BDR theory for different parameters. In Sec.~\ref{GWPinRST}, we analyze the GW polarizations in STR theory. Finally, Sec.~\ref{Conclusion} presents our conclusions and discussions.

\par Throughout this paper, the speed of light is set as $ c=1 $, and the metric signature is $ (-,+,+,+) $. The Greek alphabet indices $ (\mu, \nu, \alpha, \beta, \cdots) $ range over spacetime indices	$ (0,1,2,3) $. The Latin alphabet indices $ (i, j, k, \cdots) $ range over only spatial indices $ (1,2,3) $.

\section{Rastall theory and its generalizations}	\label{Rastall+generalizations}

\par In this section, we first briefly introduce Rastall theory~\cite{Rastall1972Generalization} and provide an overview of BDR theory~\cite{Carames:2014twa}. Then we generalize BDR theory to STR theory by considering a more general ST theory~\cite{clifton2012modified, PhysRevD.1.3209, bergmann1968comments, Maggiore:1999wm}.

\subsection{Rastall theory}

\par In Rastall theory~\cite{Rastall1972Generalization}, it is proposed that the conventional conservation law, $ \nabla^\mu T_{\mu\nu} = 0 $, is actually debatable in curved spacetime. It is assumed that the covariant divergence of energy–momentum tensor is proportional to the gradient of the curvature scalar, i.e.,
\begin{equation}	\label{Rastall_modify}
	\nabla^\mu T_{\mu\nu} = \frac{1-\lambda}{16\pi G}\nabla_\nu R,
\end{equation}
where the coefficient is expressed in this form for the sake of simplicity in the field equation. The parameter $ \lambda $ quantifies the deviation from the conservation of the energy-momentum tensor, and general relativity is recovered when $ \lambda = 1 $.

\par The modification in Eq.~\eqref{Rastall_modify} violates the Bianchi identity that the Einstein tensor satisfies. Therefore, an additional term is added to the field equation to ensure compatibility with the Bianchi identity. The field equation of Rastall theory is thus obtained as 
\begin{equation}	\label{Rastall_FEq}
	R_{\mu\nu} - \frac{\lambda}{2}g_{\mu\nu}R = 8\pi G T_{\mu\nu}  .
\end{equation}
Clearly, it reduces to the Einstein equation when $ \lambda = 1 $. 

\par Note that the trace of Eq.~\eqref{Rastall_FEq} gives 
\begin{equation}	\label{RastallFieldEq_trace}
	( 1-2\lambda )R = 8\pi G T  .
\end{equation}
It is easy to see that if $ \lambda = 1/2 $, it leads to $ T=0 $, which is not generally true. As a result, in Rastall theory, the case of $ \lambda = 1/2 $ is supposed to be excluded~\cite{Rastall1972Generalization}. However, in the following two generalized theories, we will see this restriction disappear.

\par Considering $ \lambda \neq 1/2 $, we deduce $ R=0 $ in the absence of matter source from Eq.~\eqref{RastallFieldEq_trace}. Consequently, the vacuum field equation of Rastall theory simplifies to $ R_{\mu\nu} = 0 $, aligning with that of general relativity. This suggests that the GW polarizations in Rastall theory are the same as those in general relativity, i.e., the plus and cross modes.

\subsection{Brans-Dicke-Rastall theory} 	\label{sec.BDR}

\par We start with BD theory, of which the action is given by~\cite{brans1961mach}
\begin{equation}	\label{BDAction}
	S = \frac{1}{16\pi}\int d^4x \sqrt{-g}\bigg( \phi R - \frac{\omega}{\phi}\nabla_\mu\phi\nabla^\mu\phi \bigg) + S_M( g_{\mu\nu},\psi)  .	
\end{equation} 
Here, $ \omega $ is the coupling parameter, and the action of the matter field $ S_M $ is a functional of the metric $ g_{\mu\nu} $ and the matter field $ \psi $.

\par Varying the action with respect to $ g_{\mu\nu} $ and $ \phi $ respectively, one derives the field equations:
\begin{align}	
	R_{\mu\nu} - \frac12 g_{\mu\nu}R &= \frac{\omega}{\phi^2}( \nabla_\mu\phi\nabla_\nu\phi - \frac12 g_{\mu\nu}\nabla_\lambda\phi\nabla^\lambda\phi ) + \frac{1}{\phi}( \nabla_\mu\nabla_\nu - g_{\mu\nu}\square )\phi + \frac{8\pi}{\phi}T_{\mu\nu},	\label{BDFieldEq1_Source}	\\
	(2\omega + 3) \square  \phi &= 8\pi T,	\label{BDFieldEq2_Source}
\end{align} 
where $ T $ is defined as $ T \equiv g^{\mu\nu}T_{\mu\nu} $.

\par By combining Rastall theory and BD theory, Ref.~\cite{Carames:2014twa} formulated BDR theory. In the context of BD theory, the Newtonian constant $ G $ is identified with $ 1/ \phi $. Therefore, in BDR theory, Eq.~\eqref{Rastall_modify} is transformed as
\begin{equation}	\label{BDR_modify}
	\nabla^\mu T_{\mu\nu} = \frac{1-\lambda}{16\pi}\phi\nabla_\nu R.
\end{equation}

\par Similar to Rastall theory, in order to ensure the Bianchi identity, the metric field equation of BDR theory takes the form~\cite{Carames:2014twa}
\begin{equation}	\label{BDRFieldEq1_Source}
	R_{\mu\nu} - \frac{\lambda}{2}g_{\mu\nu}R = \frac{\omega}{\phi^2}(\nabla_\mu\phi\nabla_\nu\phi - \frac12 g_{\mu\nu}\nabla_\rho\phi\nabla^\rho\phi) + \frac{1}{\phi}(\nabla_\mu\nabla_\nu - g_{\mu\nu}\square)\phi + \frac{8\pi}{\phi}T_{\mu\nu}.
\end{equation}
It is worth pointing out that Eq.~\eqref{BDRFieldEq1_Source} can be derived from an action if and only if $ \lambda = 1 $~\cite{Carames:2014twa}. However, in other geometrical frameworks, like the Weyl geometry, there may exist an action that can derive the field equations~\cite{smalley1984variational, Almeida:2013dba}.

\par The trace of Eq.~\eqref{BDRFieldEq1_Source} gives
\begin{equation} \label{BDRFieldEq1_Source_trace}
	(2\lambda-1)R = \frac{\omega}{\phi^2}\nabla_\mu\phi\nabla^\mu\phi+\frac{3}{\phi}\square\phi-\frac{8\pi}{\phi}T.
\end{equation}
It can be seen that Eq.~\eqref{BDRFieldEq1_Source_trace} cannot provide any information about the Ricci scalar $ R $ when $ \lambda = 1/2 $. Therefore, the scalar field equation and the polarization contents should be classified according to the value of $ \lambda $.

\par In the case of $ \lambda = 1/2 $, Eq.~\eqref{BDRFieldEq1_Source_trace} becomes
\begin{equation} \label{BDR_1/2_FE1_Source_trace}
	\frac{\omega}{\phi^2}\nabla_\mu\phi\nabla^\mu\phi+\frac{3}{\phi}\square\phi-\frac{8\pi}{\phi}T = 0,
\end{equation}
which can be considered as the scalar field equation.

\par In the case of $ \lambda \neq 1/2 $, the Ricci scalar $ R $ can be given by Eq.~\eqref{BDRFieldEq1_Source_trace} as 
\begin{equation}	\label{BDR_not1/2_FE1_Source_trace}
	R = \frac{1}{2\lambda-1}\bigg(\frac{\omega}{\phi^2}\nabla_\mu\phi\nabla^\mu\phi+\frac{3}{\phi}\square\phi-\frac{8\pi}{\phi}T\bigg).
\end{equation}
With this relation, one gets an equivalent form of Eq.~\eqref{BDRFieldEq1_Source}:
\begin{equation}	\label{BDR_not1/2_FE1_Source}
	\begin{aligned} 
		R_{\mu\nu} - \frac{1}{2}g_{\mu\nu}R &= \frac{\omega}{\phi^2}\bigg[\nabla_\mu\phi\nabla_\nu\phi + \frac{\lambda}{2(1-2\lambda)} g_{\mu\nu}\nabla_\rho\phi\nabla^\rho\phi\bigg]	\\
		&\quad\, + \frac{1}{\phi} \bigg[\nabla_\mu\nabla_\nu\phi + \frac{1+\lambda}{2(1-2\lambda)} g_{\mu\nu}\square\phi \bigg] + \frac{8\pi}{\phi} \bigg[T_{\mu\nu} -\frac{1-\lambda}{2(1-2\lambda)}g_{\mu\nu}T \bigg] .	
	\end{aligned}
\end{equation}
This form enables us to obtain the scalar field equation utilizing the Bianchi identity:
\begin{equation}	\label{BDR_not1/2_FE2_Source}
	\bigg[ 3\lambda-2(1-2\lambda)\omega \bigg]\square\phi = - \omega(1-\lambda)\frac{\nabla_\mu\phi\nabla^\mu\phi}{\phi} + 8\pi\lambda T.
\end{equation}

\par Consequently, the field equations of BDR theory consist of the metric field equation~\eqref{BDRFieldEq1_Source}, along with the scalar field equations~\eqref{BDR_1/2_FE1_Source_trace} for $ \lambda = 1/2 $ and~\eqref{BDR_not1/2_FE2_Source} for $ \lambda \neq 1/2 $.

\subsection{Scalar-tensor-Rastall theory}	\label{sec.RST}

\par Following the proposal of BD theory, abundant ST theories with richer structures have been developed in Refs.~\cite{clifton2012modified, PhysRevD.1.3209, bergmann1968comments, Maggiore:1999wm, faraoni2004scalar, banerjee2001cosmic}. Now, we generalize BDR theory on the basis of a more general ST theory with the action~\cite{clifton2012modified, PhysRevD.1.3209, bergmann1968comments, Maggiore:1999wm}
\begin{equation}	\label{STAction}
	S = \frac{1}{16\pi}\int d^4x \sqrt{-g}\bigg[ \phi R - \frac{\omega(\phi)}{\phi}\nabla_\mu\phi\nabla^\mu\phi + V(\phi) \bigg] + S_M( g_{\mu\nu},\psi).
\end{equation}
Compared to BD theory, the coupling parameter is generalized to a coupling function of $ \phi $, and a scalar potential $  V(\phi) $ is added to the action. Varying the action with respect to $ g_{\mu\nu} $ and $ \phi $ respectively, we obtain the field equations:
\begin{align}	
	&\begin{aligned}
		R_{\mu\nu} - \frac12 g_{\mu\nu}R 
		&= \frac{\omega(\phi)}{\phi^2}( \nabla_\mu\phi\nabla_\nu\phi - \frac12 g_{\mu\nu}\nabla_\lambda\phi\nabla^\lambda\phi ) \\
		&\quad\, + \frac{1}{\phi}( \nabla_\mu\nabla_\nu - g_{\mu\nu}\square )\phi + \frac{V(\phi)}{2\phi}g_{\mu\nu} + \frac{8\pi}{\phi}T_{\mu\nu} ,
	\end{aligned}	\label{STFieldEq1_Source}	\\
	&[ 3+2\omega(\phi) ]\square\phi + \omega'(\phi) \nabla_\mu\phi\nabla^\mu\phi + \phi V'(\phi) - 2 V(\phi)= 8\pi T.	\label{STFieldEq2_Source}
\end{align}
It can be shown that BD theory is recovered when $ \omega(\phi) $ is a constant and $ V(\phi) \rightarrow 0 $, while general relativity is recovered when $ \omega(\phi) \rightarrow \infty $, $ \omega'(\phi)/\omega^2 \rightarrow 0 $, and $ V(\phi) \rightarrow 0$~\cite{clifton2012modified}. 

\par Now, we establish STR theory by combining Rastall theory and this ST theory. Keeping Eq.~\eqref{BDR_modify} in mind and following the method used in Ref.~\cite{Carames:2014twa}, in order to ensure the Bianchi identity, the metric field equation of STR theory is obtained as
\begin{equation}	\label{RSTFieldEq1_Source}
	R_{\mu\nu} - \frac{\lambda}{2}g_{\mu\nu}R = \frac{\omega(\phi)}{\phi^2}(\nabla_\mu\phi\nabla_\nu\phi - \frac12 g_{\mu\nu}\nabla_\rho\phi\nabla^\rho\phi) + \frac{1}{\phi}(\nabla_\mu\nabla_\nu - g_{\mu\nu}\square)\phi + \frac{V(\phi)}{2\phi}g_{\mu\nu} + \frac{8\pi}{\phi} T_{\mu\nu}.
\end{equation}
Here we also have $ \nabla^\mu T_{\mu\nu} = \tfrac{1-\lambda}{16\pi}\phi\nabla_\nu R $, and ST theory is recovered when $ \lambda = 1 $.

\par The trace of Eq.~\eqref{RSTFieldEq1_Source} gives
\begin{equation}	\label{RSTFieldEq1_Source_trace}
	(2\lambda-1)R = \frac{\omega(\phi)}{\phi^2}\nabla_\rho\phi\nabla^\rho\phi + \frac{3}{\phi}\square\phi - \frac{2V(\phi)}{\phi} - \frac{8\pi}{\phi} T .
\end{equation}
Similar to BDR theory, Eq.~\eqref{RSTFieldEq1_Source_trace} gives no information about $ R $ when $ \lambda = 1/2 $. 

\par In the case of $ \lambda = 1/2 $, Eq.~\eqref{RSTFieldEq1_Source_trace} becomes
\begin{equation}	\label{RST_1/2_FE1_Source_trace}
	\frac{\omega(\phi)}{\phi^2}\nabla_\rho\phi\nabla^\rho\phi + \frac{3}{\phi}\square\phi - \frac{2V(\phi)}{\phi} - \frac{8\pi}{\phi} T = 0 ,
\end{equation}
which can be considered as the scalar field equation.

\par In the case of $ \lambda \neq 1/2 $, the Ricci scalar $ R $ is given by
\begin{equation}	\label{RST_not1/2_FE1_Source_trace}
	R = \frac{1}{2\lambda-1}\bigg[\frac{\omega(\phi)}{\phi^2}\nabla_\rho\phi\nabla^\rho\phi + \frac{3}{\phi}\square\phi - \frac{2V(\phi)}{\phi} - \frac{8\pi}{\phi} T\bigg] .
\end{equation}
With this relation, we can rewrite Eq.~\eqref{RSTFieldEq1_Source} as
\begin{equation}	\label{RST_not1/2_FE1_Source}
	\begin{aligned} 
		R_{\mu\nu} - \frac{1}{2}g_{\mu\nu}R &= \frac{\omega(\phi)}{\phi^2}\bigg[\nabla_\mu\phi\nabla_\nu\phi + \frac{\lambda}{2(1-2\lambda)} g_{\mu\nu}\nabla_\rho\phi\nabla^\rho\phi\bigg]	\\
		&\quad\, + \frac{1}{\phi} \bigg[\nabla_\mu\nabla_\nu\phi + \frac{1+\lambda}{2(1-2\lambda)} g_{\mu\nu}\square\phi \bigg]\\
		&\quad\, - \frac{1}{2(1-2\lambda)}\frac{V(\phi)}{\phi}g_{\mu\nu} 
		+ \frac{8\pi}{\phi} \bigg[T_{\mu\nu} -\frac{1-\lambda}{2(1-2\lambda)}g_{\mu\nu}T \bigg] ,	
	\end{aligned}
\end{equation}
which enables us to obtain the scalar field equation by employing the Bianchi identity:
\begin{equation}	\label{RST_not1/2_FE2_Source}
	\begin{aligned}
		\bigg[ 3\lambda-2(1-2\lambda)\omega(\phi) \bigg]\square\phi &= - \bigg[ (1-\lambda)\omega(\phi) - (1-2\lambda)\phi\omega'(\phi) \bigg]\frac{\nabla_\mu\phi\nabla^\mu\phi}{\phi} 	\\
		&\quad\, + 2\lambda V(\phi) + (1-2\lambda)\phi V'(\phi)	+ 8\pi\lambda T .
	\end{aligned}
\end{equation}

\par Consequently, the field equations of STR theory consist of the metric field equation~\eqref{RSTFieldEq1_Source}, along with the scalar field equations~\eqref{RST_1/2_FE1_Source_trace} for $ \lambda = 1/2 $ and~\eqref{RST_not1/2_FE2_Source} for $ \lambda \neq 1/2 $.

\section{Gauge invariants and polarization modes of GWs} \label{GIandPM}

\par In this section, we provide an overview of the scalar-vector-tensor decomposition of the metric perturbation, and the construction of gauge invariants~\cite{Bardeen:1980kt, Flanagan:2005yc, Caprini:2018mtu}. These methods enable us to analyze the polarization modes of GWs~\cite{Alves:2023rxs}.

\par The investigation of GW polarization modes usually involves a weak field. It is a common strategy to consider linear approximations in a Minkowski background. The dynamical variables are decomposed into terms of background and perturbation:
\begin{equation}	\label{perturbation}
	\begin{aligned}
		g_{\mu\nu} &= \eta_{\mu\nu} + h_{\mu\nu}, \quad | h_{\mu\nu} | \ll 1,	\\
		\phi &= \phi_0 + \delta\phi, \qquad | \delta\phi | \ll | \phi_0 |.
	\end{aligned}
\end{equation}
Here, $ \eta_{\mu\nu} $ is the Minkowski metric and $ \phi_0 $ is a constant background scalar field. In the following, $ \eta^{\mu\nu} $ and $ \eta_{\mu\nu} $ are used to raise and lower the indices, such as $ h^{\mu\nu} \equiv \eta^{\mu\alpha}\eta^{\nu\beta}h_{\alpha\beta} $. Accordingly, the perturbation of the inverse metric is given by $ g^{\mu\nu} = \eta^{\mu\nu} - h^{\mu\nu} $.

\subsection{Gauge invariants}

\par According to the behavior of the components of $ h_{\mu\nu} $ under spatial rotation transformations, $  h_{\mu\nu} $ can be decomposed into four scalars, two vectors, and one tensor as~\cite{Flanagan:2005yc, Alves:2023rxs}
\begin{equation}	\label{3+1Decomp.}
	\begin{aligned}	
		h_{00} &= 2\psi,	\\
		h_{0i} &= \beta_i + \partial_i \gamma,	\\
		h_{ij} &= -2\varphi\delta_{ij} + ( \partial_i \partial_j - \frac13 \delta_{ij}\nabla^2)\chi + \frac12(  \partial_i\epsilon_j + \partial_j\epsilon_i ) + h_{ij}^{\text{TT}} ,	\\
	\end{aligned}
\end{equation}
together with the constraints
\begin{equation}	\label{3+1Decomp.T}
	\partial^i \beta_i = 0,\qquad \partial^i \epsilon_i = 0,\qquad \partial^j h_{ij}^{\text{TT}} = 0,\qquad \delta^{ij} h_{ij}^{\text{TT}} = 0.
\end{equation}
This is known as the scalar-vector-tensor decomposition. The vector $ h_{0i} $ is sequentially decomposed into two parts: the transverse part and the longitudinal part; the tensor $ h_{ij} $ is sequentially decomposed into four parts: the trace part, the longitudinal traceless part, the longitudinal transverse part, and the transverse traceless part.

\par Now, the ten degrees of freedom in the symmetric tensor $ h_{\mu\nu} $ are redistributed into four degrees of freedom in scalars $ \psi, \varphi, \gamma, \chi  $, four degrees of freedom in transverse vectors $  \beta_i, \epsilon_i $, and two degrees of freedom in transverse traceless tensor $ h_{ij}^{\text{TT}} $. 

\par Let us discuss the transformation of the above scalars, vectors and tensor under the coordinate transformation
\begin{equation}	\label{CoordTran}
	x'^\mu  = x^\mu + \xi^\mu(x) ,
\end{equation}
and accordingly the gauge transformations
\begin{subequations}
	\begin{align}
		h'_{\mu\nu}(x') &=  h_{\mu\nu}(x) -  \partial_\mu\xi_\nu - \partial_\nu\xi_\mu  ,\label{GaugeTrans_ha}	 \\
		\bar h'_{\mu\nu}(x') &=  \bar h_{\mu\nu}(x) -  \partial_\mu\xi_\nu - \partial_\nu\xi_\mu + \eta_{\mu\nu}\partial_\rho\xi^\rho  ,
	\end{align}
\end{subequations}
where the definition of $ \bar h_{\mu\nu} $ is given by $ \bar h_{\mu\nu} \equiv h_{\mu\nu} - \frac12 \eta_{\mu\nu} h $, and $ h \equiv \eta^{\mu\nu} h_{\mu\nu} $ is the trace of $ h_{\mu\nu} $.

\par We parameterize $ \xi_\mu $ as~\cite{Flanagan:2005yc, Alves:2023rxs}
\begin{equation}	
	\xi_\mu = (\xi_0, \xi_i) \equiv (A, B_i + \partial_i C), \qquad \partial^i B_i = 0,
\end{equation}
where $ B_i $ and $ \partial_iC $ are the transverse and longitudinal parts of $ \xi_{\mu} $, respectively. Henceforth, the first and second derivatives with respect to time of any variable $ X $ are denoted by $ \dot X \equiv \partial_0 X $ and $  \ddot{X} \equiv - \partial^0\partial_0 X = \partial_0 \partial_0 X  $, respectively. One can thus specify the gauge transformation~\eqref{GaugeTrans_ha}:
\begin{equation}	
	\begin{aligned}	
		h'_{00} &= h_{00} - 2\dot A ,	\\
		h'_{0i} &= h_{0i} - (\partial_i A + \dot B_i + \partial_i \dot C)  ,	\\
		h'_{ij} &= h_{ij} -  (\partial_i B_j + \partial_j B_i + 2\partial_i \partial_j C) .	\\
	\end{aligned}
\end{equation}
Under this transformation, the variables transform as
\begin{equation}	\label{GaugeTrans_SVT}
	\begin{aligned}	
		\psi \quad &\rightarrow \quad \psi - \dot A ,	\\
		\gamma \quad &\rightarrow \quad \gamma - \dot C - A ,	\\
		\varphi \quad &\rightarrow \quad \varphi + \frac13 \nabla^2 C ,	\\
		\chi \quad &\rightarrow \quad \chi - 2C ,	\\
		\beta_i \quad &\rightarrow \quad \beta_i - \dot B_i ,	\\
		\epsilon_i \quad &\rightarrow \quad \epsilon_i - 2 B_i ,	\\
		h_{ij}^{\text{TT}}  \quad &\rightarrow \quad h_{ij}^{\text{TT}}.
	\end{aligned}
\end{equation}

\par Finally, according to Eq.~\eqref{GaugeTrans_SVT}, we can construct a set of gauge invariants: 
\begin{equation}	\label{GaugeInvariants}
	\begin{aligned}	
		\Psi &\equiv -\psi + \dot \gamma - \frac12 \ddot \chi,	\\
		\Phi &\equiv -\varphi - \frac16 \nabla^2 \chi,	\\
		\Xi_i &\equiv \beta_i -\frac12 \dot \epsilon_i	\qquad (\partial^i \Xi_i = 0).
	\end{aligned}
\end{equation}
Additionally, $ h_{ij}^{\text{TT}} $ itself is already a gauge invariant. Bardeen constructed the gauge invariants in a cosmological background~\cite{Bardeen:1980kt}, and Eq.~\eqref{GaugeInvariants} together with $ h_{ij}^{\text{TT}} $ are actually the Bardeen framework in a Minkowski background.

\par Now we have obtained four gauge invariants: $ \Psi $, $ \Phi $, $ \Xi_i $, and $h_{ij}^{\text{TT}} $. The degrees of freedom decrease from ten to six, containing two in the scalars $ \Psi$ and $ \Phi $, two in the transverse vector $ \Xi_i $, and two in the transverse traceless tensor $ h_{ij}^{\text{TT}} $. This implies that Eq.~\eqref{GaugeInvariants} is equivalent to a gauge choice that fixes four gauge degrees of freedom.

\subsection{Polarization modes of GWs}

\par Using the gauge invariants, one can express the polarization modes of GWs expediently. We begin with the geodesic deviation equation, which describes the relative motion of two adjacent test particles, O and P:
\begin{equation} \label{GDEq}
	\frac{\mathrm{D}^2 \xi^\mu}{\mathrm{D}\tau^2} = -R^\mu_{\alpha\nu\beta}\xi^\nu\frac{\mathrm{d} x^\alpha}{\mathrm{d}\tau}\frac{\mathrm{d} x^\beta}{\mathrm{d}\tau} .
\end{equation}
Here, $ \xi^\mu $ represents the relative displacement between O and P. Generally, the detector is situated far from the source, and moves at a nonrelativistic speed $ \mathrm{d}x^i / \mathrm{d}\tau \ll \mathrm{d}x^0 / \mathrm{d}\tau $. This allows for the consideration of the weak-field and low-speed case, where $ \mathrm{d}x^\mu / \mathrm{d}\tau = (1,0,0,0) + \mathcal{O}(h) $. The proper time and covariant derivative can be substituted by coordinate time and ordinary derivative, respectively. In the weak-field and low-speed limit, Eq.~\eqref{GDEq} simplifies to
\begin{equation} \label{GDEq_Ri0j0}
	\frac{\mathrm{d}^2 \xi^i}{\mathrm{d}t^2} = -R^i_{0j0}\xi^j .
\end{equation}

\par
\begin{figure}[h]
	\centering
	\includegraphics[width=1\textwidth]{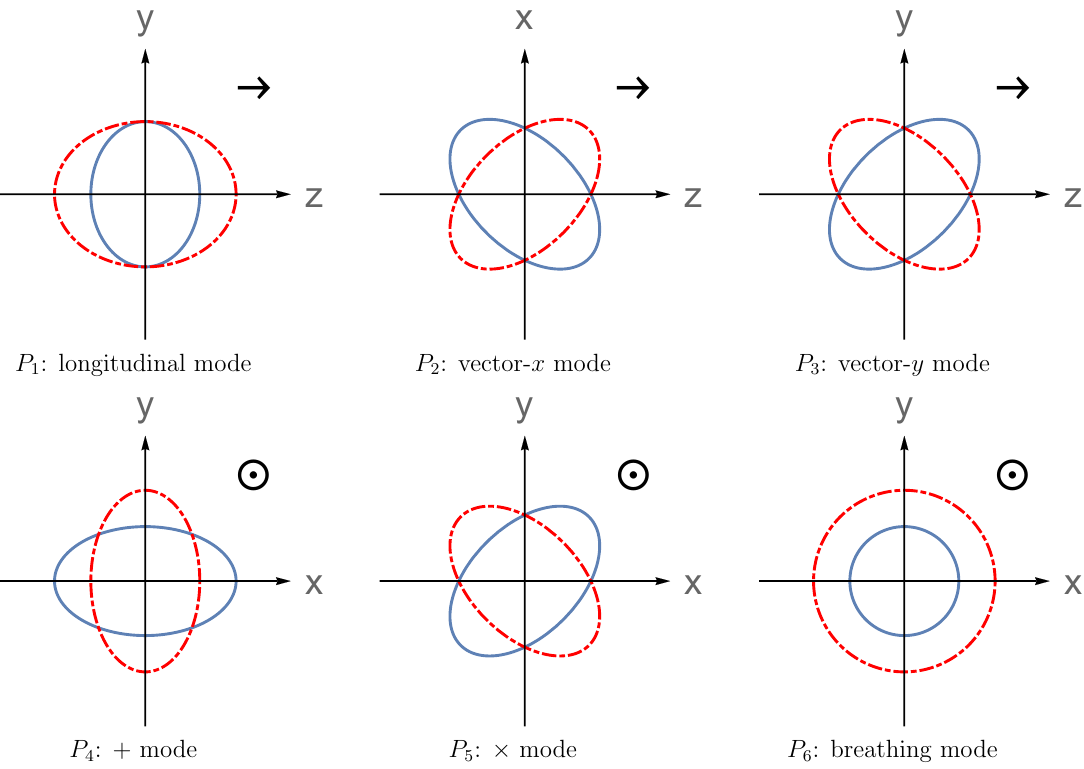}
	\caption{The displacements of freely falling test particles induced by the six polarization modes of GWs~\cite{Eardley:1973zuo}. The test particles are initially arranged on a spherical surface. The GWs propagate in the $ +z $ direction, as indicated by the symbols in the top right corner of each subplot. Both the solid and dashed lines represent the relative positions of the test particles at their maximum displacements, with a phase difference of half a period between them.}
	\label{Figure1}
\end{figure}

 From Eq.~\eqref{GDEq_Ri0j0}, we can find that the effect of GWs on the test particles is entirely determined by the components of the Riemann tensor $ R^i_{0j0} = R_{i0j0} $, which encompass all the independent components associated with GWs and induce a direct observational effect. The motion of test particles mirrors the polarization modes of GWs. Consequently, adopting the coordinate system in which the GWs propagate along the direction of $ +z $, one characterizes the polarization modes of GWs by $ R_{i0j0} $ as~\cite{Eardley:1973zuo}
\begin{equation} \label{R_i0j0_Polar.Mat.}
	R_{i0j0} = 
	\begin{bmatrix}
		\dfrac12(P_4 + P_6) & \dfrac12 P_5 & P_2 & \\
		\dfrac12 P_5 & \dfrac12(-P_4 + P_6) & P_3 &	\\
		P_2 &  P_3 & P_1 &	\\
	\end{bmatrix} .
\end{equation}
By this relation, $ P_1, \cdots, P_6 $ correspond to the longitudinal, vector-$ x $, vector-$ y $, plus, cross, and breathing modes, respectively. The displacements of freely falling test particles when the GWs with each of these modes pass them are shown in Fig. \ref{Figure1}.

\par Furthermore, under linear approximations in a Minkowski background, $  R_{i0j0} $ can be expressed in terms of gauge invariants~\cite{Alves:2023rxs}:
\begin{equation}	\label{GI_R_i0j0}
	R_{i0j0} = -\ddot\Phi \delta_{ij} + \partial_i \partial_j \Psi + \frac12 ( \partial_i\dot\Xi_j + \partial_j\dot\Xi_i ) - \frac12 \ddot h_{ij}^{\text{TT}}.
\end{equation}
Then we obtain the expression of the Ricci tensor
\begin{equation}	\label{GI_R_ab}
	\begin{aligned}	
		R_{00} &= \nabla^2\Psi - 3\ddot\Phi ,	\\
		R_{0i} &= -2\partial_i\dot\Phi - \frac12\nabla^2\Xi_i ,	\\
		R_{ij} &= - \delta_{ij}(-\ddot\Phi + \nabla^2\Phi) - \partial_i\partial_j (\Psi+\Phi) - \frac12(\partial_i\dot\Xi_j + \partial_j\dot\Xi_i) - \frac12 \square h_{ij}^{\text{TT}},
	\end{aligned}
\end{equation}
and the Ricci scalar
\begin{equation}	\label{GI_R}
	R = -2\nabla^2\Psi + 6\ddot\Phi - 4\nabla^2\Phi.
\end{equation}

\par The relation between polarization modes and gauge invariants can be given according to Eqs.~\eqref{R_i0j0_Polar.Mat.} and~\eqref{GI_R_i0j0}. Now we focus on the plane waves propagating along the direction of $ +z $. Then the gauge invariants are functions of the retarded time $ u \equiv t - z/v $ and their derivatives with respect to $ x $ or $ y $ vanish for plane waves. So the components of Eq.~\eqref{GI_R_i0j0} can be written as
\begin{equation}
	\begin{array}{lll}
 		{R_{1010} = -\ddot{\Phi} - \ddot h_+ / 2, \quad} & { 	} & {   }\\ 
 		{R_{1020} = -\ddot h_\times / 2,      } & { R_{2020} = -\ddot{\Phi} + \ddot h_+ / 2, \quad  } & { } 	\\
 		{R_{1030} =  \partial_3\dot \Xi_1 / 2, } & {R_{2030} =  \partial_3\dot \Xi_2 / 2, } & {  R_{3030} = -\ddot{\Phi} + \partial_3\partial_3 \Psi,  } \\  
 	\end{array}
\end{equation}
where $ h_+ \equiv h_{11}^{\text{TT}} = -h_{22}^{\text{TT}} $ and $ h_\times \equiv h_{12}^{\text{TT}} = h_{21}^{\text{TT}}$. Hence, considering Eq.~\eqref{R_i0j0_Polar.Mat.}, we obtain the polarization modes expressed in terms of gauge invariants:

\begin{equation}		\label{Polar_GI}
	\begin{array}{lll}
		{ P_1 = -\ddot \Phi + \partial_3\partial_3\Psi, \quad} & { P_2 = \partial_3\dot \Xi_1 /2, \quad	} & { P_3 = \partial_3\dot \Xi_2 /2, \quad }\\ 
		{ P_4 = -\ddot h_+,      } & { P_5 = -\ddot h_\times, \quad  } & { P_6 = -2\ddot \Phi. } 	\\
	\end{array}
\end{equation}

\par Clearly, the scalars $ \Psi $ and $ \Phi $ contribute to the scalar modes $ P_1 $ and $ P_6 $, the vector $ \Xi_i $ contributes to the vector modes $ P_2 $ and $ P_3 $, and the tensor $ h_{ij}^{\text{TT}} $ contributes to the tensor modes $ P_4 $ and $ P_5 $.

\par The polarization modes of GWs in BD theory and ST theory have been studied. In BD theory, in addition to the two tensor modes, there is an extra transverse breathing mode, $ P_6 $, excited by the massless scalar field~\cite{Eardley:1973zuo}. In ST theory, besides the two tensor modes, a mixture of the breathing and longitudinal modes is excited by the massive scalar field~\cite{Maggiore:1999wm, Alves:2023rxs}.

\section{Polarization modes of GWs in Brans-Dicke-Rastall theory} \label{GWPinBDR}

\par In Sec.~\ref{sec.BDR}, we have pointed out that the field equations in BDR theory are different for the parameters $ \lambda = 1/2 $ and $ \lambda \neq 1/2 $. Therefore, the polarization modes also need to be discussed separately in these two cases.

\subsection{$ \lambda = 1/2 $}

\par For convenience, here we rewrite the field equations of BDR theory with $ \lambda = 1/2 $, namely Eqs.~\eqref{BDRFieldEq1_Source} and~\eqref{BDR_1/2_FE1_Source_trace}, as follows:
\begin{align}
	&R_{\mu\nu} - \frac{\lambda}{2}g_{\mu\nu}R = \frac{\omega}{\phi^2}(\nabla_\mu\phi\nabla_\nu\phi - \frac12 g_{\mu\nu}\nabla_\rho\phi\nabla^\rho\phi) + \frac{1}{\phi}(\nabla_\mu\nabla_\nu - g_{\mu\nu}\square)\phi + \frac{8\pi}{\phi}T_{\mu\nu},	\label{BDRFieldEq1_Source_copy}	\\
	&\frac{\omega}{\phi^2}\nabla_\mu\phi\nabla^\mu\phi+\frac{3}{\phi}\square\phi-\frac{8\pi}{\phi}T = 0. \label{BDR_1/2_FE1_Source_trace_copy}
\end{align}

\par Consider the case without a source. Under the linear approximations~\eqref{perturbation}, Eq.~\eqref{BDR_1/2_FE1_Source_trace_copy} becomes
\begin{equation}	\label{BDR_1/2_2}
	\square\delta\phi = 0.
\end{equation}
With this massless Klein-Gordon equation, the linearized form of Eq.~\eqref{BDRFieldEq1_Source_copy} in vacuum is
\begin{equation}	\label{BDR_1/2_1}
	R_{\mu\nu}[h] - \frac{1}{4}\eta_{\mu\nu}R[h] =  \partial_\mu\partial_\nu\frac{\delta\phi}{\phi_0}.
\end{equation}
The notation $ [h] $ after a specific quantity represents the corresponding linearization. Equations~\eqref{BDR_1/2_1} and~\eqref{BDR_1/2_2} are the linearized field equations in vacuum with $ \lambda = 1/2 $.

\par Now, we rewrite the field equation~\eqref{BDR_1/2_1} in terms of gauge invariants. According to Eqs.~\eqref{GI_R_ab} and~\eqref{GI_R}, the $ 00 $ component of Eq.~\eqref{BDR_1/2_1} is
\begin{equation}	\label{BDR_1/2_00}
	\nabla^2\Psi - 3\ddot\Phi - 2\nabla^2\Phi = 2\frac{\delta\ddot\phi}{\phi_0},
\end{equation}
and the $ 0i $ component of Eq.~\eqref{BDR_1/2_1} is
\begin{equation}	\label{BDR_1/2_0i}
	-2\partial_i\dot\Phi - \frac12\nabla^2\Xi_i = \partial_i\frac{\delta\dot\phi}{\phi_0}.
\end{equation}
The divergence of Eq.~\eqref{BDR_1/2_0i} gives
\begin{equation}
	\nabla^2( 2\dot\Phi + \frac{\delta\dot\phi}{\phi_0} ) = 0,
\end{equation}
which indicates
\begin{equation}	\label{BDR_Phi}
	\Phi = -\frac12 \frac{\delta\phi}{\phi_0}
\end{equation}
for the solutions of propagation. Considering the scalar field equation $ \square\delta\phi = 0 $, we obtain the wave equation of $ \Phi $:
\begin{equation}	\label{BDR_PhiEq}
	\square \Phi = 0.
\end{equation}
Substituting Eq.~\eqref{BDR_Phi} to Eq.~\eqref{BDR_1/2_0i}, we get 
\begin{equation}
	\nabla^2\Xi_i = 0,
\end{equation}
which implies
\begin{equation}	\label{BDR_Xi}
	\Xi_i = 0
\end{equation}
for the solutions of propagation. Substituting Eq.~\eqref{BDR_Phi} to Eq.~\eqref{BDR_1/2_00} and using Eq.~\eqref{BDR_PhiEq}, we obtain
\begin{equation}
	\nabla^2( \Psi - \Phi ) = 0.
\end{equation}
Then we get the solution
\begin{equation}	\label{BDR_PhiPsi}
	\Psi = \Phi = -\frac12 \frac{\delta\phi}{\phi_0}.
\end{equation}
Observe that by applying Eqs.~\eqref{BDR_PhiEq}, \eqref{BDR_Xi}, and \eqref{BDR_PhiPsi}, from the $ ij $ components of Eq.~\eqref{BDR_1/2_1} we derive
\begin{equation}	\label{BDR_hTTEq}
	\square h^{TT}_{ij} = 0.
\end{equation}
Finally we conclude the field equations in terms of gauge invariants:
\begin{equation}	\label{BDR_GI}
	\Psi = \Phi = -\frac12 \frac{\delta\phi}{\phi_0},	\qquad
	\square \Phi = 0,	\qquad
	\Xi_i = 0,	\qquad
	\square h^{TT}_{ij} = 0.
\end{equation}
Equation~\eqref{BDR_GI} implies the presence of the tensor polarization modes and the massless scalar polarization mode, as well as the absence of the vector polarization modes. 

\par Assuming that the plane GWs propagate along the direction of $ +z $, due to the linearity of the wave equation \eqref{BDR_1/2_2}, we can write the plane wave solution of the scalar field equation $ \square \delta\phi = 0 $ as
\begin{equation}
	\delta\phi = \phi_1 e^{ikx}, \qquad k^\mu = (\Omega,0,0,\Omega),
\end{equation}
with $ kx \equiv k_\mu x^\mu $. Here, $ \Omega $ is the frequency of GWs, and $ \phi_1 $ is a constant amplitude. With Eqs.~\eqref{BDR_GI} and~\eqref{GI_R_i0j0}, $ R_{i0j0} $ can be expressed in the form of
\begin{equation}
	R_{i0j0} = \partial_i\partial_j\Phi - \delta_{ij} \ddot\Phi - \frac12 \ddot h^{TT}_{ij} .
\end{equation}
Eventually, we obtain the polarizations
\begin{equation}		\label{BD_1/2_Polar.}
	\begin{aligned}
		P_1 &= R_{3030} = 0,	\\
		P_2 &= R_{1030} = 0,	\\
		P_3 &= R_{2030} = 0, \\
		P_4 &= R_{1010} - R_{2020} = -\ddot h_{+},	\\
		P_5 &= 2R_{1020} = -\ddot h_{\times},	\\
		P_6 &= R_{1010} + R_{2020} = -\Omega^2\frac{\delta\phi}{\phi_0}.	\\
	\end{aligned}
\end{equation}
Consequently, BDR theory allows the plus mode, the cross mode, and the massless breathing mode in the case of $ \lambda = 1/2 $.

\subsection{$ \lambda \neq 1/2 $}

\par For the sake of argument, here we rewrite the field equations of BDR theory with $ \lambda \neq 1/2 $, namely Eqs.~\eqref{BDR_not1/2_FE1_Source} and~\eqref{BDR_not1/2_FE2_Source}, as follows:
\begin{align}
	&\begin{aligned} 
		R_{\mu\nu} - \frac{1}{2}g_{\mu\nu}R &= \frac{\omega}{\phi^2}\bigg[\nabla_\mu\phi\nabla_\nu\phi + \frac{\lambda}{2(1-2\lambda)} g_{\mu\nu}\nabla_\rho\phi\nabla^\rho\phi\bigg]	\\
		&\quad\, + \frac{1}{\phi} \bigg[\nabla_\mu\nabla_\nu\phi + \frac{1+\lambda}{2(1-2\lambda)} g_{\mu\nu}\square\phi \bigg] + \frac{8\pi}{\phi} \bigg[T_{\mu\nu} -\frac{1-\lambda}{2(1-2\lambda)}g_{\mu\nu}T \bigg] ,	
	\end{aligned}		\label{BDR_not1/2_FE1_Source_copy}		\\
	&
	\bigg[ 3\lambda-2(1-2\lambda)\omega \bigg]\square\phi =  - \omega(1-\lambda)\frac{\nabla_\mu\phi\nabla^\mu\phi}{\phi} + 8\pi\lambda T. 	\label{BDR_not1/2_FE2_Source_copy}
\end{align}

\par Consider the case without a source. Under the linear approximations~\eqref{perturbation}, Eq.~\eqref{BDR_not1/2_FE2_Source_copy} becomes
\begin{equation}	\label{BDR_not1/2_2}
	\square\delta\phi = 0.
\end{equation}
Here, we do not consider the special case of $ 3\lambda-2(1-2\lambda)\omega = 0 $. With this massless Klein-Gordon equation, the linearized form of Eq.~\eqref{BDR_not1/2_FE1_Source_copy} in vacuum is
\begin{equation}	
	G_{\mu\nu}[h] =  \partial_\mu\partial_\nu  \frac{\delta\phi}{\phi_0} ,
\end{equation}
of which the trace is $ R[h] = 0 $ due to Eq.~\eqref{BDR_not1/2_2} and the condition $ \lambda \neq 1/2 $. This equation thus simplifies to
\begin{equation}	\label{BDR_not1/2_1}
	R_{\mu\nu}[h]  =  \partial_\mu\partial_\nu  \frac{\delta\phi}{\phi_0} .
\end{equation}
Actually, the linearized field equations~\eqref{BDR_not1/2_1} and~\eqref{BDR_not1/2_2} are exactly the same as the linearized forms of field equations~\eqref{BDFieldEq1_Source} and~\eqref{BDFieldEq2_Source} in BD theory. Consequently, the polarization modes in the case of $ \lambda \neq 1/2 $ are also the same as those in BD theory, i.e., the plus mode, the cross mode, and the massless breathing mode.

\subsection{Summary}
\par From the discussions above, we arrive at the conclusion that, regardless of the value of $ \lambda $, the polarization modes of GWs in BDR theory are the plus mode, the cross mode, and the massless breathing mode
\begin{equation}		\label{BDR_Polar.}
	\begin{aligned} 
		P_4 &= -\ddot h_{+},	\\
		P_5 &= -\ddot h_{\times},	\\
		P_6 &= -\Omega^2\frac{\delta\phi}{\phi_0}.
	\end{aligned}
\end{equation}
Although the polarization modes show no difference between BDR theory and BD theory under a Minkowski background, richer properties may emerge under other backgrounds, such as a cosmological background, which deserves further exploration.

\section{Polarization modes of GWs in scalar-tensor-Rastall theory}	\label{GWPinRST}

\par In Sec.~\ref{sec.RST}, we have pointed out that the field equations in STR theory are different for $ \lambda = 1/2 $ and $ \lambda \neq 1/2 $. However, for the latter case $ \lambda \neq 1/2 $, the scenario with $ \lambda = 1 $ leads to a special result regarding polarization modes. Therefore, the polarization modes also need to be discussed separately in these cases.

\subsection{$ \lambda = 1/2 $}

\par We rewrite the field equations of STR theory with $ \lambda = 1/2 $, namely Eqs.~\eqref{RSTFieldEq1_Source} and~\eqref{RST_1/2_FE1_Source_trace}:
\begin{align}
	&\begin{aligned}
		R_{\mu\nu} - \frac{\lambda}{2}g_{\mu\nu}R 
		& = \frac{\omega(\phi)}{\phi^2}(\nabla_\mu\phi\nabla_\nu\phi - \frac12 g_{\mu\nu}\nabla_\rho\phi\nabla^\rho\phi) \\
		&\quad\, + \frac{1}{\phi}(\nabla_\mu\nabla_\nu - g_{\mu\nu}\square)\phi + \frac{V(\phi)}{2\phi}g_{\mu\nu} + \frac{8\pi}{\phi} T_{\mu\nu},
	\end{aligned}		\label{RSTFieldEq1_Source_copy}	\\
	&\frac{\omega(\phi)}{\phi^2}\nabla_\rho\phi\nabla^\rho\phi + \frac{3}{\phi}\square\phi - \frac{2V(\phi)}{\phi} - \frac{8\pi}{\phi} T = 0.	 \label{RST_1/2_FE1_Source_trace_copy}
\end{align}
For our analysis, we consider the scenario without a source. The functions $ \omega(\phi) $ and $ V(\phi) $ can be expanded in a Taylor series as
\begin{align}
	V(\phi) &= V(\phi_0) + V'(\phi_0)\delta\phi + \frac12V''(\phi_0)\delta\phi^2 + \mathcal{O}(\delta\phi^3),	\\
	\omega(\phi) &= \omega(\phi_0) + \omega'(\phi_0)\delta\phi + \frac12\omega''(\phi_0)\delta\phi^2 + \mathcal{O}(\delta\phi^3).
\end{align}
The values of $ V(\phi) $ and $ \omega(\phi) $ in a Minkowski background are $ V(\phi_0) $ and $ \omega(\phi_0) $ respectively. They are supposed to satisfy the field equations~\eqref{RSTFieldEq1_Source_copy} and~\eqref{RST_1/2_FE1_Source_trace_copy} in vacuum with $ T_{\mu\nu} = 0 $. This gives the background equations
\begin{equation}
	\begin{aligned}
		V(\phi_0)\eta_{\mu\nu}/\phi_0 &= 0,		\\
		2V(\phi_0)/\phi_0 &= 0,
	\end{aligned}
\end{equation}
which require
\begin{equation}	\label{RST_1/2_V0}
	V(\phi_0) = 0.
\end{equation}
However, the value of $ V'(\phi_0) $ is not constrained here.

\par Considering the condition~\eqref{RST_1/2_V0}, under the linear approximations~\eqref{perturbation}, we simplify Eq.~\eqref{RST_1/2_FE1_Source_trace_copy} to
\begin{equation}	\label{RST_1/2_2}
	(\square - m^2) \delta\phi = 0,\qquad m^2 \equiv \frac23V'(\phi_0).
\end{equation}
With this massive Klein-Gordon equation, the linearized form of Eq.~\eqref{RSTFieldEq1_Source_copy} in vacuum is
\begin{equation}	\label{RST_1/2_1}
	R_{\mu\nu}[h] - \frac14\eta_{\mu\nu}R[h] = (\partial_\mu\partial_\nu - \frac14\eta_{\mu\nu}\square)\frac{\delta\phi}{\phi_0}.
\end{equation}
Equations~\eqref{RST_1/2_1} and~\eqref{RST_1/2_2} are the linearized field equations in vacuum with $ \lambda = 1/2 $.

\par Now, we rewrite the field equation~\eqref{RST_1/2_1} in terms of gauge invariants. According to Eqs.~\eqref{GI_R_ab} and~\eqref{GI_R}, the $ 00 $ component of Eq.~\eqref{RST_1/2_1} is
\begin{equation}	\label{RST_1/2_00}
	2\nabla^2\Psi - 6\ddot\Phi - 4\nabla^2\Phi = 3\frac{\delta\ddot\phi}{\phi_0} + \nabla^2\frac{\delta\phi}{\phi_0},
\end{equation}
and the $ 0i $ component of Eq.~\eqref{RST_1/2_1} is
\begin{equation}	\label{RST_1/2_0i}
	-2\partial_i\dot\Phi - \frac12\nabla^2\Xi_i = \partial_i\frac{\delta\dot\phi}{\phi_0}.
\end{equation}
Similar to the calculation in Sec.~\ref{GWPinBDR}, taking the divergence of Eq.~\eqref{RST_1/2_0i} and considering the scalar field equation $ (\square - m^2)\delta\phi = 0 $, we find the wave equation
\begin{equation}	\label{RST_PhiEq}
	(\square - m^2) \Phi = 0,
\end{equation}
and the solution
\begin{equation}	\label{RST_PhiPsiXi}
	\Psi = \Phi = -\frac12 \frac{\delta\phi}{\phi_0}, \qquad \Xi_i = 0.
\end{equation}
The $ ij $ component of Eq.~\eqref{RST_1/2_1} gives
\begin{equation}	\label{RST_hTTEq}
	\square h^{TT}_{ij} = 0,
\end{equation}
where we have used Eq.~\eqref{RST_PhiPsiXi}. Finally we conclude the field equations in terms of gauge invariants as
\begin{equation}	\label{RST_GI}
	\Psi = \Phi = -\frac12 \frac{\delta\phi}{\phi_0},	\qquad
	(\square - m^2) \Phi = 0,	\qquad
	\Xi_i = 0,	\qquad
	\square h^{TT}_{ij} = 0.
\end{equation}
Equation~\eqref{RST_GI} implies the presence of the tensor polarization modes and the massive scalar polarization mode, as well as the absence of the vector polarization modes.

\par  Assuming that the GWs propagate along the direction of $ +z $, we can write the solution of the scalar field equation $ (\square - m^2)\delta\phi = 0 $:
\begin{equation}
	\delta\phi = \phi_1 e^{ikx}, \qquad k^\mu = (\Omega,0,0,\sqrt{\Omega^2 - m^2}).
\end{equation}
The wave velocity of GWs with scalar mode is thus $ v = \sqrt{\Omega^2 - m^2} / \Omega $. With Eqs.~\eqref{RST_GI} and~\eqref{GI_R_i0j0}, the components of $ R_{i0j0} $ can be expressed as 
\begin{equation}	
	\begin{aligned}
		R_{1010} &= -\ddot\Phi - \frac12 \ddot h_{11}^{TT} = -\ddot\Phi - \frac12 \ddot h_+,	\\
		R_{2020} &= -\ddot\Phi - \frac12 \ddot h_{22}^{TT} = -\ddot\Phi + \frac12 \ddot h_+,	\\
		R_{3030} &= \partial_3\partial_3\Phi - \ddot\Phi - \frac12 \ddot h_{33}^{TT} = \square\Phi = m^2\Phi,	\\
		R_{1020} &= - \frac12 \ddot h_{12}^{TT} = - \frac12 \ddot h_\times,	\\
		R_{1030} &= 0,	\\
		R_{2030} &= 0.	\\
	\end{aligned}
\end{equation}
Eventually, we obtain the polarizations:
\begin{equation}	
	\begin{aligned}
		P_1 &= R_{3030} =  -\frac12m^2\frac{\delta\phi}{\phi_0}, 	\\
		P_2 &= R_{1030} = 0 , 	\\
		P_3 &= R_{2030} = 0 , 	\\
		P_4 &= R_{1010} - R_{2020} = -\ddot h_+ , 	\\
		P_5 &= 2R_{1020} = -\ddot h_\times , 	\\
		P_6 &= R_{1010} + R_{2020} = \frac{\delta\ddot\phi}{\phi_0} = -\Omega^2 \frac{\delta\phi}{\phi_0}. 	\\
	\end{aligned}
\end{equation}
Consequently, STR theory allows the plus mode, the cross mode, and the massive mixture of the breathing and longitudinal modes in the case of $ \lambda = 1/2 $, where $ P_1 $ and $ P_6 $ correspond to a common independent variable $ \delta\phi $. This conclusion is the same as in ST theory. If $m^2 = 0$, the longitudinal mode vanishes and BD theory is recovered.

\par The absolute value of the ratio of the longitudinal mode to the breathing mode amplitude is
\begin{equation}	
	\left | \frac{P_1}{P_6} \right | = \frac{m^2}{2\Omega^2} = \frac{V'(\phi_0)}{3\Omega^2}.
\end{equation}
Considering the limitations of the observational data~\cite{chen2021non, chen2023search}, $ V'(\phi_0) $ should be close to 0.

\subsection{$ \lambda \neq 1/2 $ and $ \lambda \neq 1 $}

\par In this subsection, the discussion excludes the case of $ \lambda = 1 $, because for $ \lambda = 1 $, ST theory is recovered, while other values of $ \lambda $ present different physical contents.

\par For convenience, here we rewrite the field equations of STR theory with $ \lambda \neq 1/2 $, i.e., Eqs.~\eqref{RST_not1/2_FE1_Source} and~\eqref{RST_not1/2_FE2_Source}, as follows:
\begin{align}
	&
	\begin{aligned} 
		R_{\mu\nu} - \frac{1}{2}g_{\mu\nu}R &= \frac{\omega(\phi)}{\phi^2}\bigg[\nabla_\mu\phi\nabla_\nu\phi + \frac{\lambda}{2(1-2\lambda)} g_{\mu\nu}\nabla_\rho\phi\nabla^\rho\phi\bigg]	\\
		&\quad\, + \frac{1}{\phi} \bigg[\nabla_\mu\nabla_\nu\phi + \frac{1+\lambda}{2(1-2\lambda)} g_{\mu\nu}\square\phi \bigg]\\
		&\quad\, - \frac{1}{2(1-2\lambda)}\frac{V(\phi)}{\phi}g_{\mu\nu} 
		+ \frac{8\pi}{\phi} \bigg[T_{\mu\nu} -\frac{1-\lambda}{2(1-2\lambda)}g_{\mu\nu}T \bigg] ,	
	\end{aligned}		\label{RST_not1/2_FE1_Source_copy}		\\
	&
	\begin{aligned}
		\bigg[ 3\lambda-2(1-2\lambda)\omega(\phi) \bigg]\square\phi &= - \bigg[ (1-\lambda)\omega(\phi) - (1-2\lambda)\phi\omega'(\phi) \bigg]\frac{\nabla_\mu\phi\nabla^\mu\phi}{\phi} 	\\
		&\quad\, + 2\lambda V(\phi) + (1-2\lambda)\phi V'(\phi) + 8\pi\lambda T	.
	\end{aligned} 	\label{RST_not1/2_FE2_Source_copy}
\end{align}
Consider the case without a source. Once again we substitute the values of variables in a Minkowski background into the field equations~\eqref{RST_not1/2_FE1_Source_copy} and~\eqref{RST_not1/2_FE2_Source_copy}. This gives the background equations
\begin{equation}
	\begin{aligned}
		V(\phi_0)\eta_{\mu\nu}/\phi_0 &= 0,		\\
		2\lambda V(\phi_0) + (1-2\lambda)\phi_0 V'(\phi_0) &= 0,
	\end{aligned}
\end{equation}
which require
\begin{equation}	\label{RST_not1/2_V0}
	V(\phi_0) = V'(\phi_0) = 0.
\end{equation}
With the condition~\eqref{RST_not1/2_V0}, under the linear approximations~\eqref{perturbation}, Eqs.~\eqref{RST_not1/2_FE1_Source_copy} and~\eqref{RST_not1/2_FE2_Source_copy} can be written as linearized field equations in vacuum
\begin{align}		
	&G_{\mu\nu}[h] = [\partial_\mu\partial_\nu + \frac{1+\lambda}{2(1-2\lambda)}\eta_{\mu\nu}\square]\frac{\delta\phi}{\phi_0},	\label{RSTFieldEq1_Linear} \\
	&(\square - m^2) \delta\phi = 0,\qquad m^2 \equiv \frac{(1-2\lambda)\phi_0V''(\phi_0)}{3\lambda-2(1-2\lambda)\omega_0}.	\label{RSTFieldEq2_Linear}
\end{align}

\par Furthermore, if we write the Einstein tensor $ G_{\mu\nu}[h] $ in linearized form:
\begin{equation}
	G_{\mu\nu} = \frac12 ( \partial_\sigma\partial_\nu h^\sigma_{\ \ \mu} + \partial_\sigma\partial_\mu h^\sigma_{\ \ \nu} - \partial_\mu\partial_\nu h - \square h_{\mu\nu} - \eta_{\mu\nu}\partial_\alpha\partial_\beta h^{\alpha\beta} + \eta_{\mu\nu}\square h ),
\end{equation}
its divergence is obtained as
\begin{equation}	\label{div_EinsT.}
	\partial^\mu G_{\mu\nu} = \frac12 ( \partial_\nu \partial_\sigma\partial_\mu h^{\sigma\mu} + \square\partial^\sigma h_{\sigma\nu} - \square\partial_\nu h - \square\partial^\mu h_{\mu\nu} ) - \frac12\partial_\nu( \partial_\alpha\partial_\beta h^{\alpha\beta} - \square h ) = 0.
\end{equation}
This equation resembles the Bianchi identity. However, it is actually a conclusion derived under the premises of a Minkowski background and linear theory, and should not be understood as the linear order of the Bianchi identity.

\par Now, taking the divergence of Eq.~\eqref{RSTFieldEq1_Linear} and combining it with Eq.~\eqref{div_EinsT.}, we obtain
\begin{equation}	
	\frac{3(1-\lambda)}{2(1-2\lambda)}\partial_\nu\square\frac{\delta\phi}{\phi_0} = 0.
\end{equation}
The premise of this subsection is that $ \lambda \neq 1/2 $ and $ \lambda \neq 1 $; therefore, this equation yields 
\begin{equation}	\label{RST_massless}
	\square\delta\phi = 0.
\end{equation}
This is the wave equation derived from the linear metric field equation~\eqref{RSTFieldEq1_Linear}, which appears to contradict the linear scalar field equation~\eqref{RSTFieldEq2_Linear}. However, it necessitates considering the constraints of both equations. Therefore, a case-by-case discussion follows.

\begin{itemize}
	\item If $V''(\phi_0) = 0$, then $m^2=0$. In this case, Eqs.~\eqref{RSTFieldEq2_Linear} and~\eqref{RST_massless} are consistent, allowing for the existence of a fluctuating scalar field. Equations.~\eqref{RSTFieldEq1_Linear} and~ \eqref{RSTFieldEq2_Linear} yield the same constraint as Eqs.~\eqref{BDR_not1/2_1} and~\eqref{BDR_not1/2_2} in BDR theory. Consequently, the permissible polarization modes of GWs are the plus mode, the cross mode, and the massless breathing mode.
	\item If $V''(\phi_0) \neq 0$, then $m^2\neq0$. In this case, Eqs.~\eqref{RSTFieldEq2_Linear} and~\eqref{RST_massless} impose the constraint $ \delta\phi= 0$, indicating the absence of a fluctuating scalar field. The field equations become exactly those of Rastall theory, allowing only the plus mode and the cross mode as permissible polarization modes.
\end{itemize}

\par In fact, since the polarization modes deviating from general relativity have not been explicitly detected to date, the search for modified gravity theories containing only two tensor degrees of freedom remains an important topic~\cite{gao2020spatially,hu2021spatially}. In the second case, the polarization modes are the same as those in general relativity. Hence, investigating STR theory under these conditions is warranted.

\subsection{Summary}

\par Scalar-tensor-Rastall theory can be divided into the following three cases according to the parameter $ \lambda $. Due to the constraints of the background equations on $ V'(\phi_0) $ and $ V''(\phi_0) $, the polarization contents of each case can be more detailed.

\par \textbf{Case \uppercase\expandafter{\romannumeral1}:} $ \lambda = 1 $. In this case, STR theory reduces to ST theory, where $ V'(\phi_0) = 0 $. If $ V''(\phi_0) = 0 $, it allows the same polarization modes as those in BD theory, i.e., the plus mode, the cross mode, and the massless breathing mode; if $ V''(\phi_0) \neq 0 $, then the plus mode, the cross mode, and the mixture of the breathing and longitudinal modes are allowed.

\par \textbf{Case \uppercase\expandafter{\romannumeral2}:} $ \lambda = 1/2 $. If $ V'(\phi_0) = 0 $, then the plus mode, the cross mode, and the massless breathing mode are allowed, which are the same as in BD theory; if $ V'(\phi_0) \neq 0 $, then the plus mode, the cross mode, and the mixture of the breathing and longitudinal modes are allowed, which are the same as in ST theory.

\par \textbf{Case \uppercase\expandafter{\romannumeral3}:} $ \lambda \neq 1 $ and $ \lambda \neq 1/2 $. In this case, $ V'(\phi_0) = 0 $. If $ V''(\phi_0) = 0 $, then the plus mode, the cross mode, and the massless breathing mode are allowed, which are the same as in BD theory; if $ V''(\phi_0) \neq 0 $, then there is no fluctuating scalar field, and only the plus mode and the cross mode are allowed, which are the same as in general relativity as well as Rastall theory.

\begin{table}[htbp!]
\begin{threeparttable}
	\resizebox{1.0\columnwidth}{!}{
		\begin{tabular}{|ccccc|l|c|}
			\hline
			\multicolumn{5}{|c|}{ Theoretical parameters in STR theory  }       
			& \multicolumn{1}{c|}{\multirow{2}{*}{\makecell{ Polarization modes \\ in STR theory }}}  	& \multicolumn{1}{l|}{\multirow{2}{*}{\makecell{Theories with the \\ same polarizations }}} \\ \cline{1-5}

			\multicolumn{1}{|c|}{$ \lambda $}                      & \multicolumn{1}{c|}{$  V'(\phi_0)  $}                    & \multicolumn{1}{c|}{$  V''(\phi_0)  $}                 & \multicolumn{2}{c|}{$ m^2 $}                         & \multicolumn{1}{c|}{}                      & \multicolumn{1}{l|}{}                    \\ \hline\hline

			\multicolumn{1}{|c|}{\multirow{2}{*}{$ \lambda = 1 $}}   & \multicolumn{1}{c|}{\multirow{2}{*}{$  V'(\phi_0) = 0  $}} & \multicolumn{1}{c|}{$  V''(\phi_0) = 0  $}               & \multicolumn{1}{c|}{\multirow{2}{*}{$-\dfrac{\phi_0V''(\phi_0)}{3+2\omega_0} $}} & $ =0 $  & $ P_4, P_5, P_6 $                                        & BD, BDR                                   \\ \cline{3-3} \cline{5-7}

			\multicolumn{1}{|c|}{}                       & \multicolumn{1}{c|}{}                      & \multicolumn{1}{c|}{$ V''(\phi_0) \neq 0 $}              & \multicolumn{1}{c|}{}                    & $ \neq 0 $ & $ P_4, P_5, (P_6, P_1)$                                       & ST                                       \\ \hline

			\multicolumn{1}{|c|}{\multirow{2}{*}{$ \lambda = 1/2 $}} & \multicolumn{1}{c|}{$  V'(\phi_0) = 0  $}                  & \multicolumn{1}{c|}{\multirow{2}{*}{ Arbitrary }} & \multicolumn{1}{c|}{\multirow{2}{*}{$ \dfrac23V'(\phi_0) $}} & $ =0  $ & $ P_4, P_5, P_6 $                                        & BD, BDR                                   \\ \cline{2-2} \cline{5-7}

			\multicolumn{1}{|c|}{}                       & \multicolumn{1}{c|}{$  V'(\phi_0) \neq 0  $}                 & \multicolumn{1}{c|}{}                    & \multicolumn{1}{c|}{}                    & $ \neq 0 $ & $ P_4, P_5, (P_6, P_1)$                                       & ST                                       \\ \hline

			\multicolumn{1}{|c|}{\multirow{2}{*}{ \makecell{$ \lambda \neq 1 $  \\ $ \lambda \neq 1/2 $} }}   & \multicolumn{1}{c|}{\multirow{2}{*}{$  V'(\phi_0) = 0  $}} & \multicolumn{1}{c|}{$  V''(\phi_0) = 0  $}               & \multicolumn{1}{c|}{\multirow{2}{*}{$ \dfrac{(1-2\lambda)\phi_0V''(\phi_0)}{3\lambda-2(1-2\lambda)\omega_0} $}} & $ =0 $  & $ P_4, P_5, P_6 $                                        & BD, BDR                                   \\ \cline{3-3} \cline{5-7}

			\multicolumn{1}{|c|}{}                       & \multicolumn{1}{c|}{}                      & \multicolumn{1}{c|}{$  V''(\phi_0) \neq 0  $}              & \multicolumn{1}{c|}{}                    & $ \neq 0 $ & $ P_4, P_5 $                                        & R, GR                                   \\ \hline
		\end{tabular}
	}
	\caption{The relationship between polarization modes and parameters in STR theory. In the penultimate column, $ (P_6, P_1) $ represents the mixture of the breathing and longitudinal modes, with only one polarization degree of freedom, and $ P_4 $, $ P_5 $, $ P_6 $, and $ P_1 $ represent the plus, cross, breathing, and longitudinal modes respectively.} 	\label{RST_Table}

\end{threeparttable}
	
\end{table}

\par The main conclusions are summarized in Tab.~\ref{RST_Table}. In a Minkowski background, GWs with the scalar mode may emerge in STR theory. The mass of these GWs, denoted as $ m $, is determined by the Rastall parameter $ \lambda $ along with the first and second derivatives of the potential $  V'(\phi_0) $ and $ V''(\phi_0)  $, while the potential function $  V(\phi_0) $ itself vanishes in a Minkowski background. 

\par Additionally, it is assumed here that the coupling parameter $ \omega_0 $ is finite. If we allow $ \omega_0 \rightarrow \infty $, then this condition would also lead to $ m^2 = 0 $, and its consequences are still included in the aforementioned summary.

\par However, it is remarkable that the extra polarization modes of GWs are not necessarily emitted by actual astrophysical source in practice~\cite{Barausse:2016eii, Yunes:2011aa, Horbatsch:2011ye}, even if these modes are predicted in a theory. For example, in ST theory, if the scalar field is globally trivial throughout spacetime, binary black holes generally do not emit GWs with scalar polarization modes, whereas binaries involving at least one neutron star do~\cite{Barausse:2016eii}. This discrepancy arises from the vanishing trace $ T $ of the energy-momentum tensor for black holes and the nonvanishing trace $ T $ for neutron stars, with the latter serving as the source of scalar-polarized GWs. The similar reasoning and results apply to the case of STR theory with $ \lambda = 1/2 $. Furthermore, binary black holes with different masses are able to emit GWs with scalar polarization modes if the scalar field, whose mass is small enough, varies slowly in time~\cite{Horbatsch:2011ye}. Additionally, in Einstein-Scalar-Gauss-Bonnet theory with certain scalar Gauss-Bonnet coupling function, GWs with scalar polarization modes can also be emitted by binary black holes without a nontrivial background scalar field~\cite{East:2021bqk, Corman:2022xqg}. Thus, the presence of GWs with scalar polarization modes depends not only on the astrophysical source but also the modified gravity theory in question~\cite{Barausse:2016eii}.

\section{Conclusion}		\label{Conclusion}

\par In this paper, after reviewing Rastall theory~\cite{Rastall1972Generalization} and BDR theory~\cite{Carames:2014twa}, we established STR theory as a generalization of BDR theory. Then, we gave a brief introduction to the gauge invariants used for analyzing the polarization modes of GWs. Subsequently, under a Minkowski background, we adopted linear approximations and obtained the linear perturbation equations of these theories. Finally, we investigated the polarization modes of GWs in BDR theory and STR theory using the gauge invariants.

\par We found that in BDR theory, the field equations take different forms for $ \lambda = 1/2 $ and $ \lambda \neq 1/2 $, but the polarization modes of GWs are the same in both cases. For any parameter $ \lambda $ in BDR theory, it allows the plus mode, the cross mode, and the breathing mode. However, the polarization contents become richer when it comes to STR theory. The field equations of this theory can be divided into three cases according to the parameter $ \lambda $, and the polarization modes for each case can be further divided into two scenarios. The polarization modes may consist of the plus mode, the cross mode, and the breathing mode. Alternatively, they could represent the plus mode, the cross mode, and the mixture of the breathing and longitudinal modes. It is also possible that only the plus mode and the cross mode exist. The specific scenario depends on the parameter $ \lambda $ and the derivatives of the potential function in a Minkowski background. The detailed relationships between the polarization modes and the parameters were summarized in Tab.~\ref{RST_Table}.

\par The work in this paper was carried out only in the context of a Minkowski background. As a simple yet practical example, the Minkowski spacetime provides straightforward constraints that simplify the complex field equations. However, if we extend our scope to other backgrounds, such as a cosmological background, and consider the perfect fluid as the background matter field, the situation becomes more complex. The nonconservation of the energy-momentum tensor~\eqref{Rastall_modify} in Rastall theory could potentially lead to more significant changes in the field equations. This suggests that further research is needed to explore whether the theories involved in this paper would exhibit different polarization modes in other possible backgrounds.

\par The detection of GWs is crucial for testing modified gravity theories. At the current stage, the means of detecting polarization modes of GWs include not only ground-based GW detectors, such as LIGO, Virgo, and KAGRA, but also ways like PTAs. The ground-based GW detectors are primarily sensitive to high-frequency GWs in the frequency range of $ 10 \sim 10^4 $ Hz, while PTAs mainly detect low-frequency GWs in the frequency range of $ 10^{-10} \sim 10^{-6} $ Hz. In recent years, a weak evidence for scalar transverse modes has been discovered in the NANOGrav data set~\cite{chen2021non, chen2023search}. However, we still need more precise detectors to confirm the existence of additional polarization modes, which would help select gravity theories and constrain theoretical parameters. This work is expected to be completed in future space-borne GW detection projects, such as LISA~\cite{LISA}, Taiji~\cite{Taiji}, and TianQin~\cite{Tianqin}. The space GW antennas mainly detect mid-frequency GWs in the range of $ 10^{-4} \sim 10 $ Hz, complementing the detections such as ground-based GW detectors and PTAs.

\section*{Acknowledgments}
We would like to express our sincere gratitude to the anonymous referee for his valuable comments and suggestions, which significantly improved the original manuscript. We also thank Yu-Peng Zhang and Chun-Chun Zhu for useful discussions. This work is supported in part by the National Key Research and Development Program of China (Grant No. 2020YFC2201503), the National Natural Science Foundation of China (Grants No. 12475056, No.123B2074, and No. 12247101), the 111 Project (Grant No. B20063), and the Department of Education of Gansu Province: Outstanding Graduate “Innovation Star” Project (Grant No. 2025CXZX-153).

\normalem
\bibliography{myref.bib}

\end{document}